\documentclass[referee]{mn2e}
\usepackage{graphicx}

\title[The Hubble flow around the Local Group]
{The Hubble flow around the Local Group}

\author[I. D. Karachentsev et al.]
{I.~D.~Karachentsev$^{1}$\thanks{E-mail:ikar@luna.sao.ru},
O. G. Kashibadze$^{1}$, D. I. Makarov$^{1}$ and R. B. Tully$^{2}$\\
$^{1}$Special Astrophysical Observatory of the Russian Academy
 of Sciences, Nizhnij Arkhyz, KChR, 369167, Russia \\
$^{2}$Institute for Astronomy of Hawaii, 2680 Woodlawn Drive,
 Honolulu, HI 96822, USA}
\begin{document}


\pagerange{\pageref{firstpage}--\pageref{lastpage}} \pubyear{2008}

\def\LaTeX{L\kern-.36em\raise.3ex\hbox{a}\kern-.15em
    T\kern-.1667em\lower.7ex\hbox{E}\kern-.125emX}

\newtheorem{theorem}{Theorem}[section]

\label{firstpage}

\maketitle

\begin{abstract}
We use updated data on distances and velocities of galaxies in the proximity of
the Local Group (LG) in order to establish properties of the local Hubble flow. For 30 neighbouring galaxies with distances $0.7 < D_{LG} <
3.0$ Mpc, the Local flow is characterized by the Hubble parameter $H_{loc} = (78\pm2)$ km s$^{-1}$ Mpc$^{-1}$, the mean-square peculiar velocity 
$\sigma_v = 25$ km s$^{-1}$, corrected for errors of radial velocity measurements ($\sim4$ km s$^{-1}$) and distance measurements ($\sim10$ km 
s$^{-1}$), as well as the radius of the zero-velocity surface $R_0 = (0.96\pm0.03)$ Mpc. The minimum value for $\sigma_v$ is achieved when the barycenter of 
the LG is located at the distance $D_c = (0.55\pm0.05) D_{M31}$ towards M31 corresponding to the Milky Way-to-M31 mass ratio $M_{MW}/M_{M31} 
\simeq 4/5$. In the reference frame of the 30 galaxies at $0.7-3.0$~Mpc, the LG barycenter has a small peculiar velocity $\sim(24\pm4)$ km s$^{-1}$ towards the Sculptor constellation. 
The derived value of $R_0$ corresponds to the total mass $M_T(LG) = (1.9\pm0.2) 10^{12} M_{\odot}$ with $\Omega_m = 0.24$ and a topologically flat universe, a value in good 
agreement with the sum of virial mass estimates for the Milky Way and M31.
\end{abstract}


\section{Introduction}

Studies of the Hubble flow of galaxies in the neighbourhood of the Local Group (LG) has been out of favour with 
cosmologists. The main reason has been a lack of reliable data on distances to even the nearest galaxies outside the LG. Early on, Lynden-Bell
(1981), Sandage (1986, 1987) and Giraud (1986, 1990) noticed that a behaviour of the Hubble flow at distances $D\sim(1-3)$ Mpc allows one to find 
the total mass $M_T$ of the LG independently of virial mass estimates for the Milky Way (MW) and Andromeda galaxy (M31) based on the motions of their companions.
A decelerating influence of the LG on nearby galaxies causes the Hubble regression $\langle V\mid D\rangle$ to cross the line of
zero velocity at a non-zero distance $R_0$. In the case of spherical symmetry with $\Lambda = 0$, the radius of the zero-velocity surface $R_0$ can be expressed via the LG mass
and the age of the Universe $T_0$ by a simple relation
$$
M_T(LG)=(\pi^2/8G)\cdot  R^3_0\cdot T_0^{-2}, \eqno(1)
$$
where G is the gravitational constant. Sandage et al. (1972) noticed another important feature of the local Hubble flow; that the dispersion of radial 
velocities of the nearby galaxies with respect to the Hubble regression is small at $\sigma_v\simeq70$ km s$^{-1}$. Later Karachentsev \&  
Makarov (2001), Ekholm et al. (2001) and Karachentsev et al. (2002) showed that nearby single galaxies and centers of nearby groups have even smaller
peculiar velocities  $\sigma_v\simeq 30$ km s$^{-1}$. As Governato et al. (1997) and Maccio et al. (2005) noticed, this low value of
$\sigma_v$ disagrees with results of N-body simulation of the velocity field of galaxies in the standard cosmological model where the density of matter is close to the critical value
$(\Omega_m\simeq1)$. Baryshev et al. (2001) and Chernin (2001) proposed to explain the low value of $\sigma_v$ by the presence of "dark energy" 
(cosmological vacuum) in the Universe, which effectively cools down the chaotic motions of galaxies. Numerical simulations, performed by Chernin et 
al. (2004) with the vacuum acting as an universal antigravity are in good correspondence with observational data of a cold local Hubble 
flow. This agreement began to be considered as evidence of the dark energy manifestation not only on the cosmological, but also on local 
scales $\sim3$ Mpc (Teerikorpi et al. 2005, Chernin 2008). However, using N-body simulations with high spatial 
resolution, Hoffman et al. (2008) have recently shown that galaxy groups like the LG with cold Hubble flow can be obtained in cosmological models with different parameters. The observed
value $\sigma_v\simeq50$ km s$^{-1}$ is not a unique characteristic of a single cosmological model.

Last year, a number of papers appeared (Hoffman et al. 2008, Peirani \& Pacheco 2008, Tikhonov \&  Klypin 2008) with interpretations of the observational data on the local Hubble flow,
but procedures in each case were not transparent.
Below we present an up-to-date list of distances and radial velocities for the 30 nearest galaxies around the LG
and use them to estimate the basic parameters that characterize the local Hubble flow.

\section{Observational data on the neighbouring galaxy flow}

Usage of the wide-field survey of the Andromeda neighbourhood (Ibata et al. 2001, Irwin et al. 2005, Ferguson et al. 2002) and of the Sloan survey archive
led to the discovery of a couple dozen new companions to the MW and M31, doubling the total number of identified dwarf galaxies in the LG. Nowadays
there are 47 members in the LG with known distances and radial velocities which permits improved derivations of the virial masses of the MW and M31.

In the close proximity of the LG there are 30 more galaxies with distances from the group centroid within $0.7<D_{LG}<3.0$ Mpc. Their main 
observational properties are listed in Table 1. The lower boundary was accepted to be 0.7 Mpc in order to exclude companions to the MW and M31 not participating in the cosmological expansion. The upper boundary was adopted to be 3.0 Mpc in order to exclude members of the nearby groups around IC342, M81 and Centaurus A, with their centers located at 3.3, 3.6 and 3.7 Mpc from us, respectively. Approximately the same distance range was used by 
Karachentsev et al. (2002) and Karachentsev \& Kashibadze (2006) while finding $R_0$ and $\sigma_v$ parameters.

In the last years, new galaxies have also been discovered in the volume just beyond the LG, but at a much lower rate than in the close proximity to the MW and M31. Improving galaxy distances based on the luminosity of the tip of red giant branch (TRGB) with respect to the distances from Karachentsev et al. (2002)
leads 6 galaxies: SDIG, NGC~247, UGCA~92, NGC~1569, UGC~8638, and ESO~383-87 to be placed beyond $D_{LG}=3.0$ Mpc. 
Other galaxies within the volume (ESO~410-005, HIZSS003, UGC~4879) have their radial velocities and/or distances newly measured.  
The improved observational data motivated us to reconsider the properties of the local Hubble flow.

In Table 1, the first and second columns are the galaxy name and its equatorial coordinates at the J2000.0 equinox, respectively. Columns (3,4)
are heliocentric velocity with its error and the velocity with respect to the LG centroid. With one exception all the velocities $V_h$ are taken from 
the NED database and their transformation into $V_{LG}$ is performed with the standard apex parameters (Karachentsev \& Makarov 1996) adopted in NED. Columns
(5) and (6) represent the distance of the galaxy from the MW (with error) and from the centroid of the LG under the assumption that it is located 0.43 Mpc away from the MW in the 
direction toward M31 (see below). Columns (7) and (8) give blue absolute magnitude and morphological type in the digital scale of the de Vaucouleurs
et al. (1976) catalogue. Column (9) contains the so-called ``tidal index'' of the galaxy according to CNG (Karachentsev et al. 2004). A negative value for 
$TI$ corresponds to an isolated galaxy of the general field while a case with $TI>0$ means the galaxy is in a zone of significant gravitational 
influence from its neighbours. Columns (10) and (11) are peculiar radial velocities of the galaxy with respect to two lines of Hubble regression shown in the upper and bottom panels of Figure 1 (see below). Finally, the last
column contains the reference to the source of data on the galaxy distance and comments.
\begin{table*}
\footnotesize
\caption{Galaxies around the Local Group with distances  $0.7 < D_{LG} < 3.0$ Mpc.}
\begin{tabular}{|l|c|r|r|r|l|l|r|r|r|r|l|} \hline
 Galaxy &  RA (J2000.0)D  &$V_{hel} \pm e$& $V_{LG}$& $D_{MW} \pm e$  &$D_{0.55}$&  $M_B$ &  $T$ &  $TI$ & $V_{pec}$ & $V_{pec'}$ & Ref.   \\
	&                 &km s$^{-1}$  & km s$^{-1}$&  Mpc     &Mpc  &  mag &    &     & km s$^{-1}$ & km s$^{-1}$ &        \\
\hline
  (1)   &       (2)       & (3)  &  (4)&  (5)     & (6) &  (7) & (8)&  (9)& (10) & (11) & (12)  \\
\hline
WLM     & 000158.1$-$152740 &-122 2& -16 & 0.97 .02 &0.82 &$-$14.06&  9 &$ $0.3   &$-$4   &$ $17   & R07    \\
AndXVIII& 000214.5$+$450520 &      &     & 1.36 .09 &0.94 &$-$ 8.7 & $-$3 &$ $0.7 &$ $    &$ $     & M08    \\
NGC55   & 001508.5$-$391313 & 129 3& 111 & 2.17 .11 &2.14 &$-$18.47&  8 &$-$0.4   &$-$16  &$ $2    & T06    \\
E410-005& 001531.4$-$321048 &  38 3&  55 & 1.94 .11 &1.86 &$-$11.60& $-$1 &$ $0.4 &$-$48  &$-$29   & T06    \\
E294-010& 002633.3$-$415120 & 107 5&  71 & 1.96 .12 &1.95 &$-$10.95& $-$1 &$ $1.0 &$-$39  &$-$20   & T06    \\
NGC300  & 005453.5$-$374057 & 144 5& 114 & 2.15 .10 &2.11 &$-$17.92&  7 &$-$0.3   &$-$10  &$ $6    & CNG    \\
NGC404  & 010926.9$+$354303 & -48 9& 195 & 3.24 .18 &2.82 &$-$16.61& $-$1 &$-$1.0 &$ $8   &$-$7    & CNG+   \\
HIZSS003& 070029.3$-$041230 & 280 1& 101 & 1.69 .17 &1.79 &$-$12.60& 10 &$-$0.6   &$ $6   &$ $2    & S05    \\
UGC4879 & 091602.2$+$525024 & -20 7&  38 & 1.10 .10 &1.08 &$-$11.50& 10 &$-$1.4   &$ $17  &$-$1    & KTF08  \\
LeoA    & 095926.4$+$304447 &  24 4& -40 & 0.80 .04 &0.96 &$-$11.68& 10 &$ $0.2   &$-$45  &$-$59   & D02    \\
SexB    & 100000.1$+$051956 & 301 1& 111 & 1.44 .06 &1.70 &$-$14.08& 10 &$-$0.7   &$ $26  &$ $14   & T06    \\
NGC3109 & 100307.2$-$260936 & 403 1& 110 & 1.34 .06 &1.70 &$-$15.70&  9 &$-$0.1   &$ $25  &$ $24   & T06    \\
Antlia  & 100404.0$-$271955 & 362 {\bf 1}&  66 & 1.25 .09 &1.61 &$-$ 9.63& 10 &$ $1.6   &$-$11  &$-$10   & T06    \\
SexA    & 101100.8$-$044134 & 324 1&  94 & 1.32 .07 &1.63 &$-$13.95& 10 &$-$0.6   &$ $16  &$ $8    & CNG    \\
DDO99   & 115053.0$+$385250 & 242 1& 248 & 2.64 .14 &2.74 &$-$13.52& 10 &$-$0.5   &$ $71  &$ $42   & CNG    \\
IC3104  & 121846.1$-$794334 & 430 5& 171 & 2.27 .19 &2.61 &$-$14.85&  9 &$-$0.5   &$ $5   &$ $17   & CNG    \\
DDO125  & 122741.8$+$432938 & 195 4& 240 & 2.74 .14 &2.81 &$-$14.32&  9 &$-$0.9   &$ $57  &$ $27   & T06    \\
GR8     & 125840.4$+$141303 & 214 3& 136 & 2.13 .11 &2.39 &$-$12.14& 10 &$-$1.2   &$-$12  &$-$30  & T06    \\
UGC8508 & 133044.4$+$545436 &  62 5& 186 & 2.69 .14 &2.67 &$-$13.09& 10 &$-$1.0   &$ $14  &$-$14   & T06    \\
KKH86   & 135433.6$+$041435 & 287 3& 209 & 2.60 .19 &2.90 &$-$10.30& 10 &$-$1.5   &$ $20  &$ $1    & T06    \\
KK230   & 140710.7$+$350337 &  62 2& 126 & 2.14 .12 &2.26 &$-$ 9.80& 10 &$-$1.0   &$-$11  &$-$32   & T06    \\
DDO187  & 141556.5$+$230319 & 152 4& 172 & 2.24 .12 &2.43 &$-$12.47& 10 &$-$1.3   &$ $21  &$ $1    & T06    \\
DDO190  & 142443.5$+$443133 & 150 3& 263 & 2.80 .14 &2.84 &$-$14.19& 10 &$-$1.3   &$ $77  &$ $51   & T06    \\
KKR25   & 161347.6$+$542216 &-139 2&  68 & 1.86 .12 &1.79 &$-$ 9.94& $-$1 &$-$0.7 &$-$28  &$-$46   & CNG+   \\
IC4662  & 174706.3$-$643825 & 308 4& 145 & 2.44 .19 &2.74 &$-$15.56&  9 &$-$0.6   &$-$31  &$-$20   & K06    \\
SagDIG  & 192959.0$-$174041 & -79 1&  21 & 1.04 .07 &1.14 &$-$11.49& 10 &$-$0.3   &$-$6   &$ $11   & CNG    \\
DDO210  & 204651.8$-$125053 &-141 2&   9 & 0.94 .04 &0.93 &$-$11.09& 10 &$-$0.1   &$ $7   &$ $26   & CNG    \\
IC5152  & 220241.9$-$511743 & 124 3&  75 & 1.97 .12 &2.08 &$-$15.56&  9 &$-$1.1   &$-$46  &$-$25   & T06    \\
Tucana  & 224149.0$-$642512 & 132 5&  11 & 0.88 .04 &1.09 &$-$ 9.16& $-$2 &$-$0.1 &$-$10  &$ $18   & CNG    \\
UA438   & 232627.5$-$322326 &  62 5&  99 & 2.22 .12 &2.15 &$-$12.92& 10 &$-$0.7   &$-$29  &$-$13   & T06    \\
KKH98   & 234534.0$+$384304 &-137 3& 151 & 2.45 .13 &2.04 &$-$10.78& 10 &$-$0.7   &$ $31  &$ $22   & CNG    \\
\hline
\multicolumn{1}{|l}{R07:}  &
\multicolumn{11}{|l|}{Rizzi L., Tully R.B., Makarov D., et al., 2007, ApJ, 661, 815} \\
\multicolumn{1}{|l}{M08:}   &
\multicolumn{11}{|l|}{McConnachie A.W., Huxor A., Martin N.F., et al. 2008, astro-ph/0806.3988}\\
\multicolumn{1}{|l}{T06:}    &
\multicolumn{11}{|l|}{Tully R.B., Rizzi L., Dolphin A.E., et al. 2006, AJ, 132, 729 (T06)}\\
\multicolumn{1}{|l}{S05:}     &
\multicolumn{11}{|l|}{Silva D.R., Massey P., DeGioia-Eastwood K., Henning P.A., 2005, ApJ, 623, 148}\\
\multicolumn{1}{|l}{KTF08:}    &
\multicolumn{11}{|l|}{Kopylov A.I., Tikhonov N.A., Fabrika S., et al. 2008, MNRAS, 387L, 45}\\
\multicolumn{1}{|l}{D02:}       &
\multicolumn{11}{|l|}{Dolphin A.E., Saha A., Claver J., et al., 2002, AJ, 123, 3154}          \\
\multicolumn{1}{|l}{K06:} &
\multicolumn{11}{|l|}{Karachentsev I.D., Dolphin A.E., Tully R.B., et al. 2006, AJ, 131, 1361} \\
\hline
\end{tabular}
\end{table*}

Most of the galaxies in the close proximity to the LG have new, more
accurate distance estimates with respect to ones given in CNG. Some cases
deserve additional comments.

{\em ESO 410-05}. This dwarf spheroidal galaxy is a companion to the spiral NGC 55. Koribalski et al. (2009) found
from an HI line observation its neutral hydrogen mass to be
$\sim10^6M_{\odot}$ and its radial velocity $V_h=+38$ km s$^{-1}$.

{\em NGC~404}. An isolated lenticular galaxy with an extended $HI$-envelope (Rio et al. 2004). We adopted its distance to be 3.26 Mpc as an average over 
three estimates made by Tonry et al. (2001), Karachentsev et al. (2002) and Tikhonov et al. (2003).

{\em UGC 4879}. The distance to this relatively bright ($B=13.78^m)$ dIr-type galaxy was found by Kopylov et al. (2008) via TRGB. The authors give its optical radial velocity to be $V_h=-70$ km s$^{-1}$. But we would rather use another estimate, $V_h=-20$ km s$^{-1}$, obtained from the 21-cm-line observations by Oosterloo (2008).

{\em KKR25}. This isolated dSph-type galaxy was considered detected by Huchtmeier et al. (2003) in the $HI$-line with Effelsberg radiotelescope with $V_h=-139$ km s$^{-1}$. However Begum \& Chengalur (2005) repeated $HI$-observations with aperture synthesis on GMRT and failed to
find any $HI$-flux from KKR25. The previous signal was probably due to extended emission of Galactic hydrogen in the foreground.

{\em And XVIII}. This dSph-type system was recently discovered by McConnachie et al. (2008) who found its distance to be 1.36 Mpc according to TRGB. 
The radial velocity for this galaxy has not been measured yet. Being separated by $\sim600$ kpc from M31, And XVIII is  probably a peripheric companion to M31 
on the opposite side from the MW.

The distribution of galaxies by distances and radial velocities with respect to the LG centroid is given at the upper panel of Fig.1. Each galaxy is 
represented by a filled circle with horizontal and vertical bars indicating the errors of measurements of distance and radial velocity of the
galaxy. For the sake of completeness we also present 47 members of the LG. They fill a vertical region with $D_{LG}<0.7$ Mpc. An inclined dotted 
line represents a linear relation with the global Hubble parameter $H_0=73$ km s$^{-1}$ Mpc$^{-1}$. The solid line corresponds to the 
regression $\langle V_{LG}|D_{LG}\rangle$ for the canonical Lemetre-Tolman (LM) model
\setcounter{equation}{1}
\begin{equation}
D = \frac{\pi^{2/3}}{2}R_0 \times  \left\{
\begin{array}{cc}
\frac{1-\cos\theta}{(\theta-\sin\theta)^{2/3}}, &
 \theta < 0\\
\frac{\cosh\theta-1}{(\sinh\theta-\theta)^{2/3}}, &
\theta > 0
\end{array}
\right.
\end{equation}

\setcounter{equation}{2}
\begin{equation}
V = \frac{\pi^{2/3}}{2}R_0T_0^{-1} \times  \left\{
\begin{array}{cc}
\frac{\sin\theta(\theta-\sin\theta)^{1/3}}{1-\cos\theta}, &
 \theta < 0\\
\frac{\sinh\theta(\sin h\theta-\theta)^{1/3}}{\cosh\theta-1}. &
\theta > 0
\end{array}
\right.
\end{equation}

The regression line drawn for $\Omega_m=0$, $T_0=13.4$ Gyr crosses the zero-velocity level at $R_0=0.91$ Mpc.

In the upper panel of Fig.2 we represent the distribution of 30 galaxies from Table 1 on the sky in equatorial coordinates. The galaxies are shown as 
circles with their sizes indicating the galaxy distance while numbers and colour reflect peculiar velocity. The irregular band  of strong Galactic extinction is indicated. The nearest groups: IC342, M81, CenA, and
Canes Venatici I cloud (CVnI), the local mini-attractors, are shown as large ellipses.
The data demonstrates little visible correlation between peculiar velocities of the galaxies and their location with respect to the neighboring groups. One can note, though, that three galaxies with the largest positive velocities (DDO99, DDO125 and DDO190) are situated in the direction 
of CVnI and probably take part in the primordial collapse of this diffuse system. In general, the distribution of $V_{pec}$ demonstrates a minor dipole effect, with an excess 
of negative peculiar velocities in the southern hemisphere.

\section{Radius of the zero-velocity surface $R_0$}

An estimate of $R_0$ for the LG was obtained by Karachentsev \& Makarov (2001) with use of the radial velocities for 20 galaxies having distances 
from 0.7 to 3.0 Mpc. The resulting value $R_0=0.96\pm0.05$ Mpc was derived with $H_0=70$ km s$^{-1}$ Mpc$^{-1}$. Later with addition of new data on 
nearby galaxies Karachentsev et al. (2002) confirmed the previous estimate by obtaining $R_0=0.94\pm0.10$ Mpc with $H_0=72$ km s$^{-1}$ Mpc$^{-1}$. A more detailed analysis of factors affecting the estimate of $R_0$ was done by Karachentsev \& Kashibadze (2006). Parameters of the local Hubble flow depend on a choice of the LG
barycenter position. In the paper cited above it was shown that the minimum value
for the sum of squares of the peculiar velocities with respect to the Hubble regression line is achieved when the distance from the MW to the barycenter is $D_c=(0.55\pm0.05)\cdot D_{M31}$
in the direction toward M31 with the adopted distance to M31 of 0.78 Mpc. Assuming $D_c/D_{M31}=0.55$ and $H_0=72$ km s$^{-1}$ Mpc$^{-1}$, Karachentsev \& Kashibadze (2006) obtained the zero-velocity radius to be $0.96\pm0.03$ Mpc and  the mean-square value of the peculiar radial velocities to be $\sigma_v=24$ km s$^{-1}$. Also, their modeling
of the local Hubble flow with possible chaotic tangential velocities of the galaxies around the LG showed that
typical tangential velocities with amplitude of 35 and 70 km s$^{-1}$ produce a statistical uncertainty in $R_0$ as $\pm0.02$ and $\pm0.04$ Mpc,respectively. 

Improved observational data on the galaxies around the LG given in Table 1 are used by us to make a new estimate of $R_0$ and its uncertainty. We 
consider a local value of the Hubble parameter $H_{loc}$ and the distance to the LG barycenter $D_c$ as arbitrary parameters. We also accept that the local peculiar velocity field
might have a dipole anisotropy with an arbitrary amplitude and direction. The dependence of 
$\sigma_v$ on the distance to the barycenter $D_c$ is shown in Fig.3. The requirement of minimum value for $\sigma_v$ yields us the following six 
parameters: $H_{loc}=(78\pm2)  km s^{-1} Mpc^{-1}, D_c/D_{M31}=0.55\pm0.05$, the dipole amplitude (in the LG frame) $V_d=(24\pm4)$ km s$^{-1}$ 
directed to RA=$336\pm34^{\circ}, \delta=-64\pm10^{\circ}$ and the radius $R_0=(0.96\pm0.03)$ Mpc.

The Hubble diagram corresponding to these parameters is persented in the bottom panel of Fig.1, and the distribution of the residual (peculiar) velocities of the galaxies 
are shown in the bottom panel of Fig.2 (their numerical values are also given in Table 1).

Despite the fact that some of the galaxies shifted positions on the Hubble diagram, the value of $R_0$ remains almost unchanged. The reason for
the stability of the quantity $R_0$ is mostly due to the fact that 6 galaxies in the vicinity of $R_0$: WLM, Leo~A, DDO~210, 
UGC~4879, Tucana, and SagDIG have the average distance $\langle D_{LG}\rangle =(0.98\pm0.05)$ Mpc and the mean radial velocity $\langle V_{LG}\rangle 
=(+4\pm11)$ km s$^{-1}$, fixing the value $R_0$ with good precision.

\section{Peculiar velocity pattern around the Local Group}

It follows from the data above that the regular Hubble flow begins from the distance $D\simeq R_0$, at the boundary of the LG. In the distance range of 
 $D_{LG}=0.7-3.0$ Mpc this flow is characterized by the local Hubble parameter $H_{loc}$ which only $(7\pm5$)\% larger than the commonly accepted global Hubble parameter $H_0=(73\pm3)  km s^{-1} Mpc^{-1}$ (Spergel et al. 2007). Expected differences between $H_{loc}$ and $H_0$ for the case of a canonical LT-model for the LG and for the 
case of a modified LT-model with $\Lambda$-term were discussed by Peirani \& Pacheco (2006, 2008).

The LG barycenter position at $D_c/D_{M31}=0.55\pm0.05$ in the direction to M31 indicates that the mass ratio of the two main LG 
members is close to unity and probably lies within a range of $M_{MW}/M_{M31}=[2/3 - 1]$. According to Fukugita \& Peebles (2004) the amplitude of internal 
rotation for the MW is $V_m(MW)=241\pm13$ km s$^{-1}$ and for M31 it is $V_m(M31)=259\pm5$ km s$^{-1}$. Since masses of
spiral galaxies are proportional to the power 3 to 4 of $V_m$, the ratio obtained for $M_{MW}/M_{M31}$ agrees fairly well with observational data on internal motions of the galaxies.

As it was noted above, the peculiar velocity field around LG has a small dipole component $V_d=24\pm4$ km s$^{-1}$ with respect to the reference frame of the LG centroid.
The direction of the dipole in equatorial ($RA=336^{\circ}, D=-64^{\circ})$, galactic ($l=325^{\circ}, b=-46^{\circ}$) or 
Supergalactic ($L=227^{\circ}, B= 1^{\circ}$) coordinates is not directed toward any neighbour group as a local mini-attractor.
However one can note that the anti-pole is very close to M81. The positive motions
in that sector could be due to streaming away from us toward IC342/M81/CVn.
On the opposite side of the sky, there are a lot of galaxies with negative
peculiar velocities but they lie essentially in two groupings: a nearer
grouping around WLM - SagDIG and a more distant grouping around NGC 55.
Both could be attracted to the nearest big mass which is us.  Then the
apparent dipole would be caused by galaxies being pulled away from us on
one side of the sky and toward us on the other. The interpretation is ambiguous
because, although there are 30 galaxies, they are clumped into only 3 or 4
independent groupings.
In the bottom panel of Fig.2 we denote the location of the Virgo cluster and the direction opposite to the Local
Void (AntiLV) by two large ellipses. Location on the sky of these two massive attractors is not associated with the local dipole. A behaviour of the local dipole as a function of the radius of the sphere around the LG was analyzed by Kashibadze (2008). By varying the radius from 2 to 6 Mpc, the dipole amplitude varies within a cap of 32 km s$^{-1}$ while its direction chaotically drifts over the sky. Therefore, the value of $V_d$ obtained can be considered as a small random value with no clear dynamical sense.

The distribution of the peculiar velocities of the galaxies after accounting for the dipole is presented in Fig.4 as a function of $D_{LG}$. Here, the inclined bars
correspond to the 1-$\sigma$ errors of distances to galaxies.

For 30 galaxies within $D_{LG}=0.7-3.0$ the mean-square dispersion of the velocities is 27 km s$^{-1}$. After reduction for the mean error of the
radial velocity measurement (4 km s$^{-1}$) and for the mean error of the distance measurement (10 km s$^{-1}$) the residual (cosmic) dispersion 
drops to $\sigma^c_v=25$ km s$^{-1}$. Hence, the peculiar motion of the LG centroid with respect to nearby galaxies (24 km s$^{-1}$) has an amplitude 
comparable with $\sigma_v$.

The local field of peculiar velocities gives us an unique opportunity to test whether the dispersion of the velocities depends on the luminosities of 
the galaxies or on the density of their environment. Upper and bottom panels
of Fig.5 represent variations of the modulus of peculiar velocity with the blue
absolute magnitude $M_B$ of the galaxy and with its tidal index $TI$, respectively. The linear regressions are shown as dashed lines. A slight tendency in the increasing of the $\mid V_{pec}\mid$ from giant galaxies to dwarf ones is seen. Whiting (2005, 2006) studied data on galaxies within 10 Mpc, and found no correlation between luminosity and peculiar velocity at all. Also, the galaxies that reside in the close proximity with others: Antlia, E294-010, E410-005, LeoA and WLM tend to have peculiar velocities just a bit surpassing those of isolated galaxies. The large peculiar velocity of KKR25 is most likely an artefact (due to the Galactic hydrogen confusion) and three dIr systems in front of the CVnI cloud (DDO~99/125/190) might have common acceleration towards the cloud center.

For some cosmological problems it is enough to know the total dispersion of peculiar velocities inside a fixed volume without distinguishing between virial motions and the motions of galaxies external to a virialized group. By summarizing the data for 30 galaxies considered above 
with data on 47 companions to the MW and M31 we obtain the peculiar radial velocity dispersion to be $\sigma^{tot}_v=79$ km s$^{-1}$ within $D_{LG}<3$ Mpc.

According to Jing et al. (1998), Branchini et al. (2001), Zehavi et al. (2002), Feldman et al. (2003) and others, the radial velocity difference for a galaxy and its nearest neighbour as a function of
their spatial separation can be a useful tool for mapping the matter distribution on scales of $\sim 1$ Mpc. 
Implementation of this method has been limited until now due to the poverty of observational data on distances to galaxies. For the more extended region of the Local Volume extending to 10 Mpc, the relation between the difference of radial velocities $\mid V_1-V_2\mid$ and spatial separations $\mid R_1 - R_2 \mid$ was shown in CNG (Karachentsev et al. 2004). The behaviour of the median difference $\mid V_1-V_2\mid$ vs. $\mid R_1 - R_2 \mid$ 
demonstrates approximately the same value $\mid V_1-V_2\mid\simeq100$ km s$^{-1}$ within $\mid R_1 - R_2 \mid<1$ Mpc and systematic increase up to $\sim250$ km 
s$^{-1}$ with increasing of the spatial separations to 3 Mpc. For the 77 nearest galaxies with known radial velocities and distances within 3 Mpc this relation is shown in Fig.6.
Here the LG members (N = 47) and the galaxies around the LG (N = 30) are denoted as open and filled circles, respectively. Although there are not so many galaxies considered, the precision of their distances (typically $\sim0.12$ Mpc) is much higher than for more distant CNG galaxies. In general the features are seen to be the same as in the CNG sample: a virial region with $\mid V_1-V_2\mid\simeq100$ km s$^{-1}$ for $\mid R_1 - R_2 \mid<1$ Mpc and than a smooth increase of the pairwise velocity differences with increase of separations because of the contribution of the Hubble component. 
To a first approximation, the kinematics of the LG and its neighbourhood looks to be typical of the more extended Local Volume.

\section{The total mass of the Local Group}

In the standard flat cosmological model with $\Lambda$-term and $\Omega_m$ as an matter component it takes a form
$$
 M_T =(\pi^2/8G)\cdot R_0^3\cdot H^2_0/f^2(\Omega_m), \eqno(4)$$
\noindent where 
$$
f(\Omega_m)=(1-\Omega_m)^{-1}-(\Omega_m/2)\cdot(1-\Omega_m)^{-3/2}\cdot
arccos h[(2/\Omega_m)-1].\eqno(5)
$$

\noindent Assuming $H_0=73$ km s$^{-1}$ Mpc$^{-1}$ that corresponds to $T_0=13.7$ Gyr with $\Omega_m=0.24$ (Spergel et al. 2007), one can rewrite (4) as
$$
(M_T/M_{\odot})_{0.24}=2.12\cdot10^{12}(R_0/Mpc)^3.\eqno(6)
$$
Therefore, for the radius of the zero-velocity surface as $R_0=0.96\pm0.03$ Mpc, the total mass of the LG turns out to be
$M_T=(1.88\pm0.18)\cdot10^{12}M_{\odot}$. Let us compare this value with estimates of the LG mass from the orbital motions of the MW and M31 companions.

The list of properties of the nearest groups of galaxies (Karachentsev 2005) yields 
$M(MW)=0.94\cdot10^{12}M_{\odot}$ and $M(M31)=0.84\cdot10^{12}M_{\odot}$ as an average of virial and orbital estimates for the subsystems around the MW and M31. Later estimates of the 
masses for the subsystems with newly found dwarf galaxies taken into account give $M(M31)=0.71\cdot10^{12}M_{\odot}$ (Geehan et al. 2006), 
$M(M31)=1.1\cdot10^{12}M_{\odot}$ (Tempel et al. 2007), $M(MW)=(1.1\pm0.2)\cdot10^{12}M_{\odot}$ (Xue et al. 2008), all being not so different from previous estimates. The total mass of the LG, found by using the motions of satellites, lies within the range $M(MW+M31)=(1.6 -2.2)\cdot10^{12}M_{\odot}$,
which is consistent with the total mass of the LG from the motions of the 30 surrounding galaxies.

It should be noted, however, that the LG mass estimate based on so-called ``timing argument'' (Kahn \&Woltjer 1959) gives a much higher value for the total
mass. This approach assumes the convergent motions of the MW and M31 as a pair without influence from any other galaxies. In order to explain the
relative approach of the MW and M31 with the observed velocity of $-$123 km s$^{-1}$, the total mass of the LG should be 
$\sim5\cdot10^{12}M_{\odot}$. However if one takes into account the radial velocities and luminosities of M33 and LMC, the most massive companions to M31 
and the MW, then the relative velocity between barycenters of MW+LMC and M31+M33 would be ($-$109), not ($-$123) km s$^{-1}$. This correction decreases
the estimate of the ``timing mass'' from $5.0\cdot10^{12}M_{\odot}$ to $3.9\cdot10^{12}M_{\odot}$. The detailed analysis performed by Li \&  White  
(2008) as well as van der Marel \& Guhathakurta (2008) shows that with the requirement of Keplerian orbits for the MW and M31 the total mass of the LG 
should be greater than $M_{min}(LG)=1.7\cdot10^{12}M_{\odot}$. Apparently, the value of the total mass $M(LG)=(1.7-2.1)\cdot10^{12}M_{\odot}$
is a reasonable consistent compromise between all three independent methods used to derive the total mass of the LG.

\section{Concluding remarks}

The recent data on radial velocities and distances for nearby galaxies demonstrate the regularity of the Hubble flow around the LG on scales of 1-3 Mpc around the LG center. The mean-square velocity of chaotic motions along the line-of-sight is about $\sigma_v=25$ km s$^{-1}$ after 
reduction for uncertainties of the radial velocity and distance measurements. Approximately the same low amplitude peculiar motions are seen around other neighbour groups (Karachentsev \& Kashibadze 2006). The dispersion of radial velocities of the centers of these groups is also of the same order, $\sim25$ km s$^{-1}$, (Karachentsev 2005) and is about 3 times lower than the typical dispersion of virial motions inside the groups. Finally, the peculiar velocity of the LG itself with respect to nearby galaxies is found to be about 24 km s$^{-1}$.

According to Peebles (1980), a typical amplitude of peculiar velocity $V_{pec}$ on a scale of $H_0r$ relates to the average density of matter $\Omega_m$ and the deviation in density from the mean $\delta$ as
$$
V_{pec}=(H_0r/3)\cdot\Omega^{0.6}_m\delta(1+\mid\delta\mid)^{-1/4}, \eqno(7)
$$
where the last term takes nonlinear correction into account (Yahil 1985). According to Karachentsev et al. (2004) and Karachentsev \& Kutkin (2005), the density excess (found by the luminosity in K-band) in the sphere of 3 Mpc radius around the MW is $\delta=4.5\pm0.5$. Assuming $\Omega_m=0.24$ one can find 
$V_{pec}/H_0r=0.416$ on the scale considered. For $H_0=73$ km s$^{-1}$ Mpc$^{-1}$ and $r=3$ Mpc the expected 3D and 1D peculiar velocities are 91 km 
s$^{-1}$ and  53 km s$^{-1}$, respectively. The observed chaotic motions of the galaxies along the line-of-sight (25 km s$^{-1}$) are half the 
expected ones (53 km s$^{-1}$).

Systematic searches for new nearby galaxies and measuring their distances and velocities allowed us to find the radius of the zero-velocity surface for the LG with an accuracy of
$\sim0.03$ Mpc, which leads to uncertainty of the total mass of the LG within 1 Mpc of around 10\%. This high precision is due to three main 
factors: 1) the reliability of measuring individual distances via the TRGB method with typical error $\sim7$\% (Rizzi et al. 2007); 2) the ``coldness'' of the local Hubble flow with a typical ``thermal'' velocity $\sigma_v=25$ km s$^{-1}$ on the scales of 1-3 Mpc; 3) a large enough number of galaxies in the distance region of (1--3)$R_0$.

Among 77 galaxies with known radial velocities inside the sphere of 3 Mpc radius, 30 of them (i.e. 39\%) are located outside the ``hot'' virial region of the LG. This ratio may be important for an understanding of cosmic structure
formation on the scale of 1 -- 3 Mpc if it turns to be typical for other groups as well. However, our knowledge of the specific number of galaxies outside the LG virialized region (39\%) may be strongly affected by observational selection. Recent special programs to search for new companions to the MW and M31 that doubled the number of known LG members (McConnachie et al. 2008) cover a small fraction of the whole sky and are not suited for systematic discovery of faint dwarf galaxies outside the LG. The example of KKR25 at the distance of 1.8 Mpc from us (Karachentsev et al. 2001) proves that galaxies with very low luminosity ($-9.9^m$) 
and old population do exist far outside $R_0$. Moreover, Pasquali et al. (2005) reported the discovery by chance of another dSph galaxy, APPLES1, with an absolute magnitude of $-8.3^m$ at a distance of 8.5 Mpc. The limiting magnitude of the SDSS survey is insufficient for finding faint dwarf 
galaxies by resolving them into RGB stars even at distances of 2--3 Mpc. The number of the APPLES1-type objects could be significant. For a rough estimate let us assume that the HST archive has about 1000 sufficiently exposed frames with ACS randomly distributed over the whole sky. With the field of
view of ACS taken into account ($\sim$ 10 sq. arcmin), to recover of order unity APPLES1-like galaxies from the whole collection, their total number should be $\sim$ 15000 in the volume with $D\leq10$ Mpc, and about 400 in the shell with $1<D<3$. This is one order higher than the number of currently known galaxies (N=30) in the same shell. On the other hand, a survey of the M81 group performed with 3.5-m CFHT telescope 
and MegaCam over a 65 sq. deg. area (Chiboukas et al. 2008) resulted in the discovery of 22 new dwarf members similar to APPLES1 in this group. However there was not a single foreground object with $D<3$ Mpc. By comparing the sky area studied with the whole sky area, one can obtain an upper limit to the 
total number of these kinds of objects within $D<3$ Mpc as $N<600$, which does not contradict the previous estimate. Next generation deep large-scale surveys of the Northern and Southern hemispheres (Ivezic et al. 2008, PanStarr,
http://pan-starr.ifa.hawaii.edu) might lead to the discovery of 
hundreds of faint dwarf galaxies in the region (1--3)$R_0$ around the LG, making the study of the local Hubble flow a very important branch of observational cosmology. 

It should be stressed that finding $R_0$ with high precision allows one to determine the total mass of the LG with uncertainty $\sim10$\%. There is consistency
of the mass estimates given by establishing the zero-velocity surface around a group and from studies of the virial motions of galaxies within the group.

{\bf Acknowledgements}

This work was supported by RFBF grants 07--02--00005, RFBF-DFG 06--02--04017 and HST grant for the programms
GO 10905 and 11126.

{}

\renewcommand{\baselinestretch}{1.2}
\newpage

\begin{figure}
\includegraphics[scale=0.5]{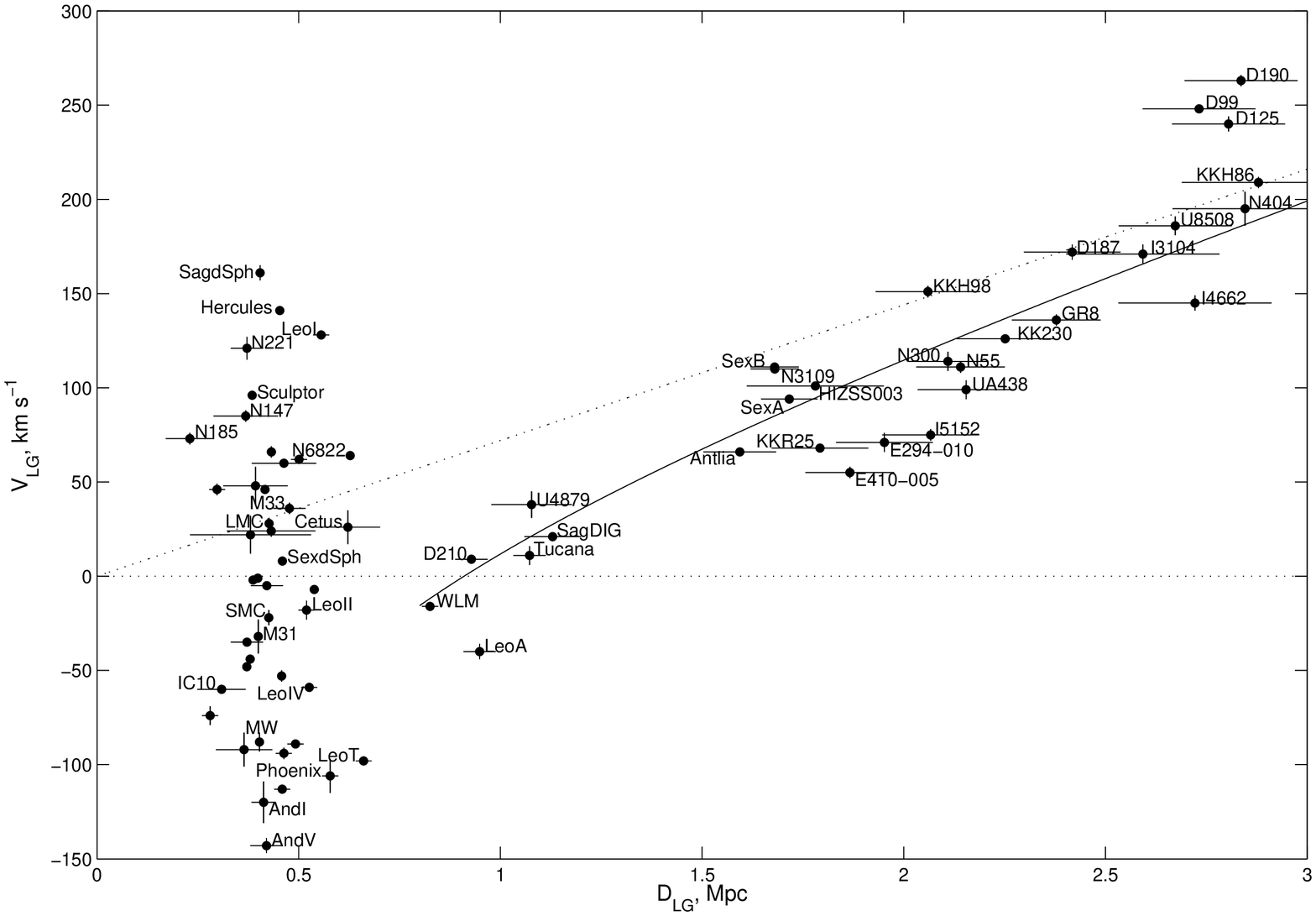}
\includegraphics[scale=0.5]{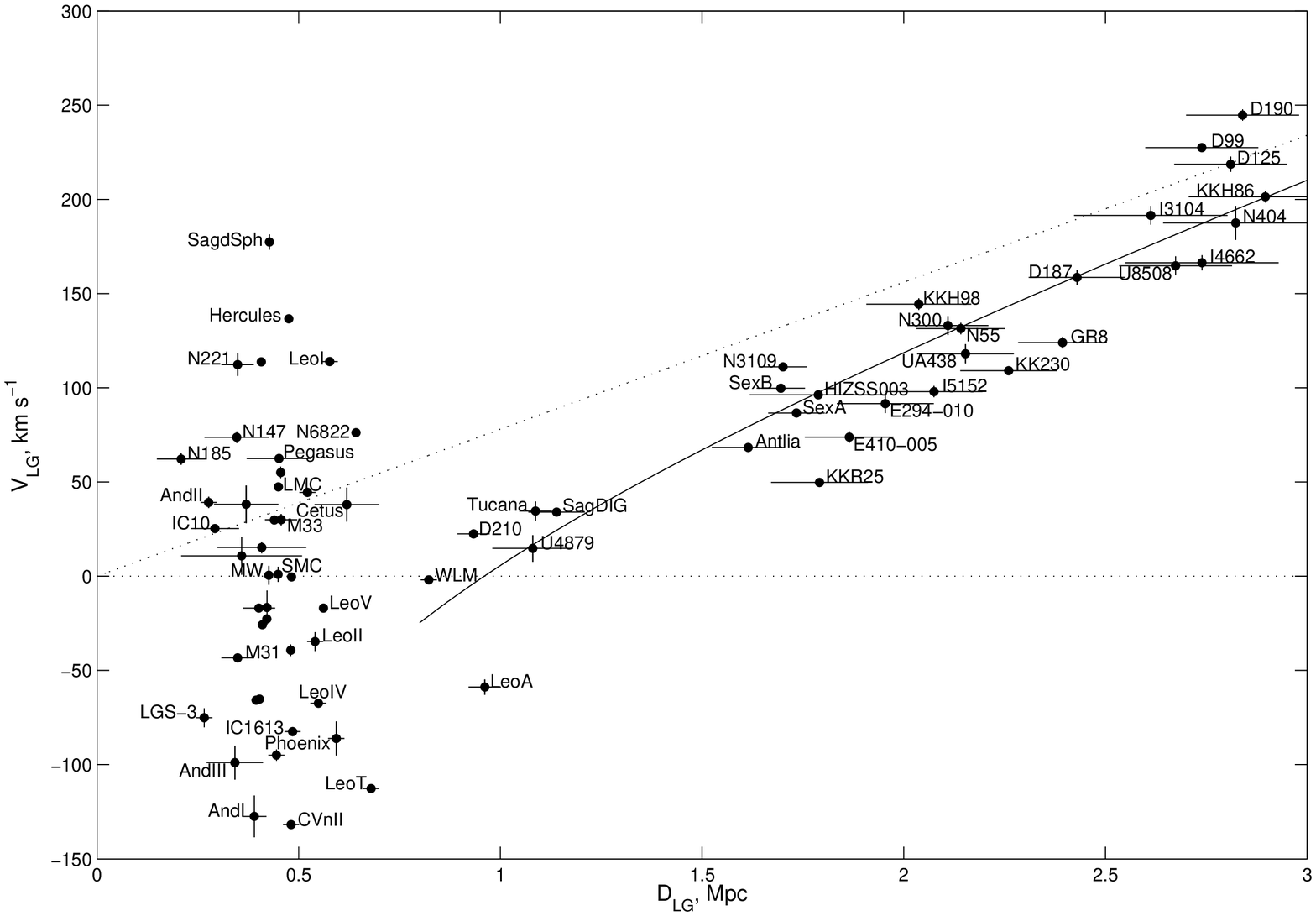}
\caption{Radial velocities and distances to galaxies in the neighbourhood of the Local Group (LG) with respect to its centroid. Horizontal and vertical bars represent the errors of distance and velocity measurements, respectively. {\em Upper}: Heliocentric velocities of galaxies $V_h$ are converted into the
LG frame, $(V_{LG}$), with the standard apex parameters: $V_A=316$ km s$^{-1}$,
$l_A=93^{\circ}$, $b_A=-4^{\circ}$ adopted in NED. {\em Bottom}: $V_h$ converted to $V_{LG}$ with an additional dipole component ($V_d=24 $ km s$^{-1}, l_d=336^{\circ}, 
b_d=-64^{\circ}$). In both the cases, the LG barycenter is situated at the distance $D_c=0.55\cdot D_{M31}$ directed toward M31. The inclined dotted line represents the 
linear Hubble relation with $H_0=73$ km s$^{-1}$ Mpc$^{-1}$ for upper panel and $H_0=78$ km s$^{-1}$ Mpc$^{-1}$ for bottom panel. Solid line is a 
non-linear regression line ((2), (3)) curved by a decelerating influence of the LG. The intersection of this line with the zero-velocity line yields $R_0$.}
\end{figure}

\begin{figure}
\includegraphics[scale=0.5,angle=-90]{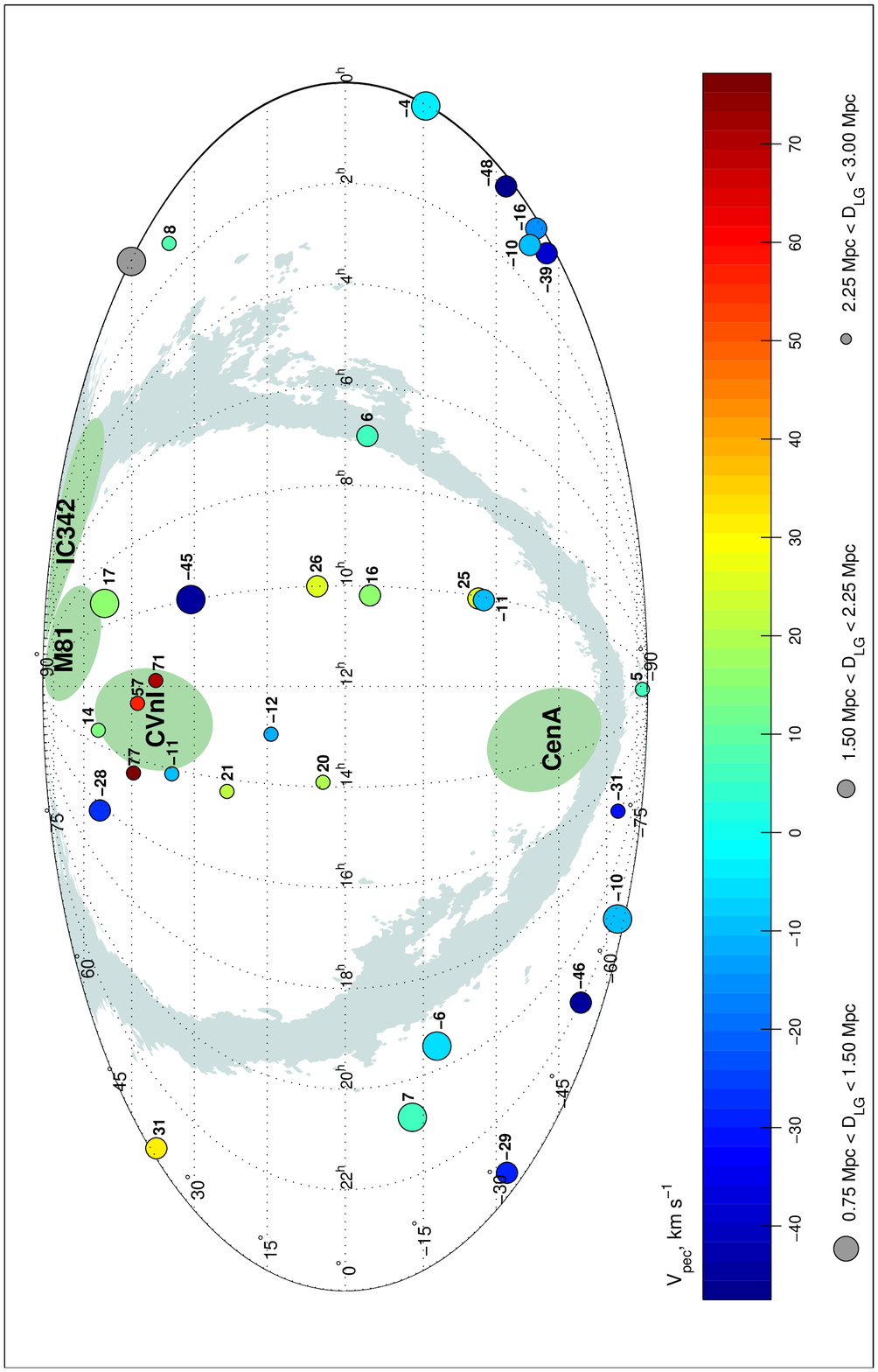}
\includegraphics[scale=0.5,angle=-90]{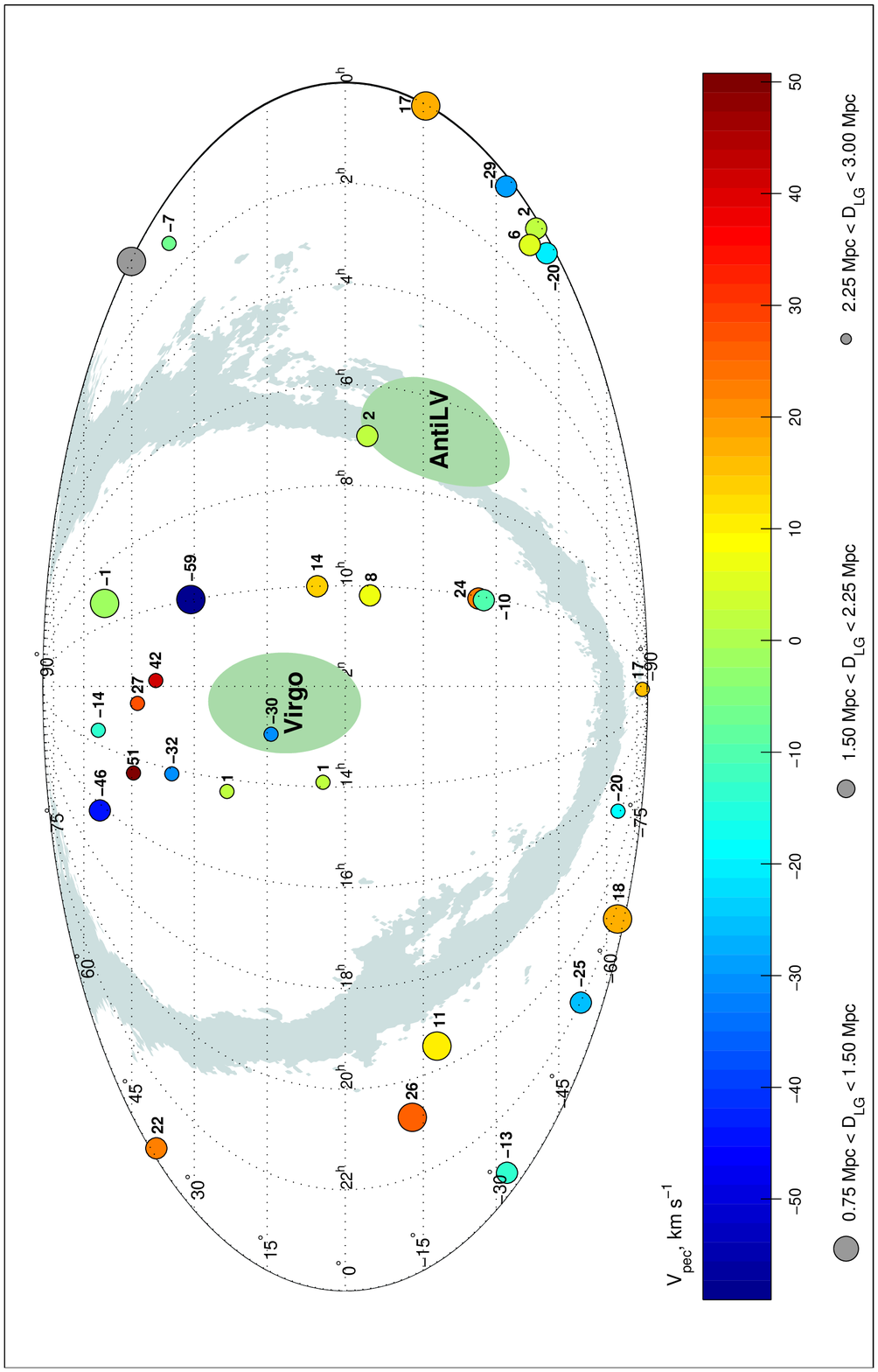}
\caption{Distribution of the 30 nearest galaxies outside the LG over the sky. Size of the circles reflects the distance to the galaxy while the number indicates the peculiar velocity of the galaxy with respect to Hubble regression in km s$^{-1}$ units. Upper and bottom panels correspond to the 
regression parameters from Fig.1 (upper) and (bottom), respectively. The positions of four closest groups, the Virgo cluster and the direction opposite to the 
Local Void are indicated with large ellipses. The fluffy band is a region with strong absorption within the MW.}
\end{figure}

\begin{figure}
\includegraphics[scale=0.5]{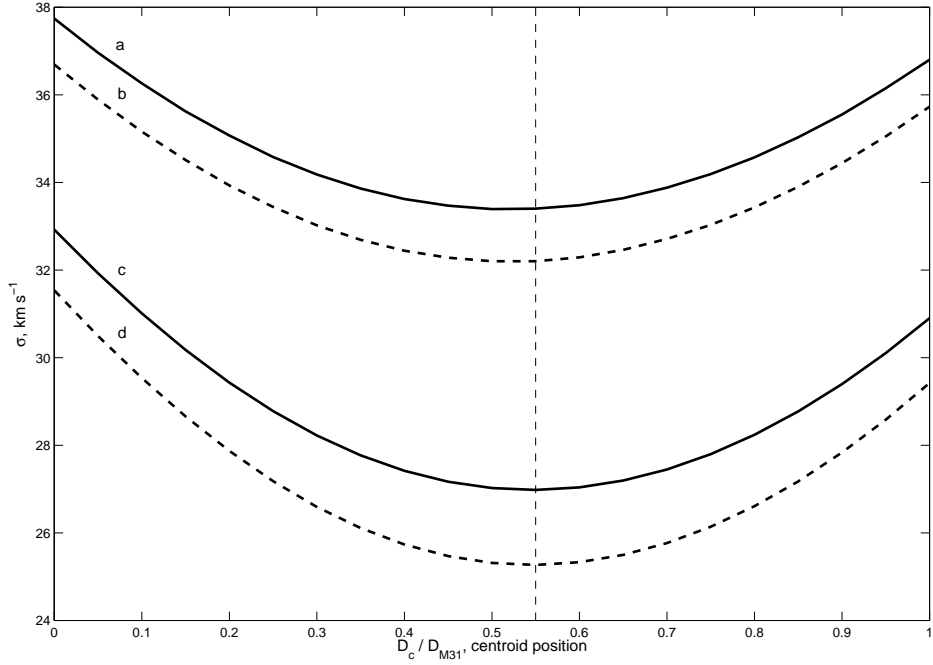}
\caption{Mean-square peculiar velocity of 30 galaxies around the LG for different positions of the LG barycenter.
a) the standard apex, $H_0=73$ km s$^{-1}$ Mpc$^{-1}$, no corrections for distance measurement error; b) the same but with the distance measurement errors taken into account; c)  $H_{loc}= 78$ km s$^{-1}$ Mpc$^{-1}$ with the dipole motion of $V_d=24$ km s$^{-1}$ taken into account; d) the same $H_{loc}$ and $V_d$ but after correcting the distance measurement errors}.
\end{figure}

\begin{figure}
\includegraphics[scale=0.5]{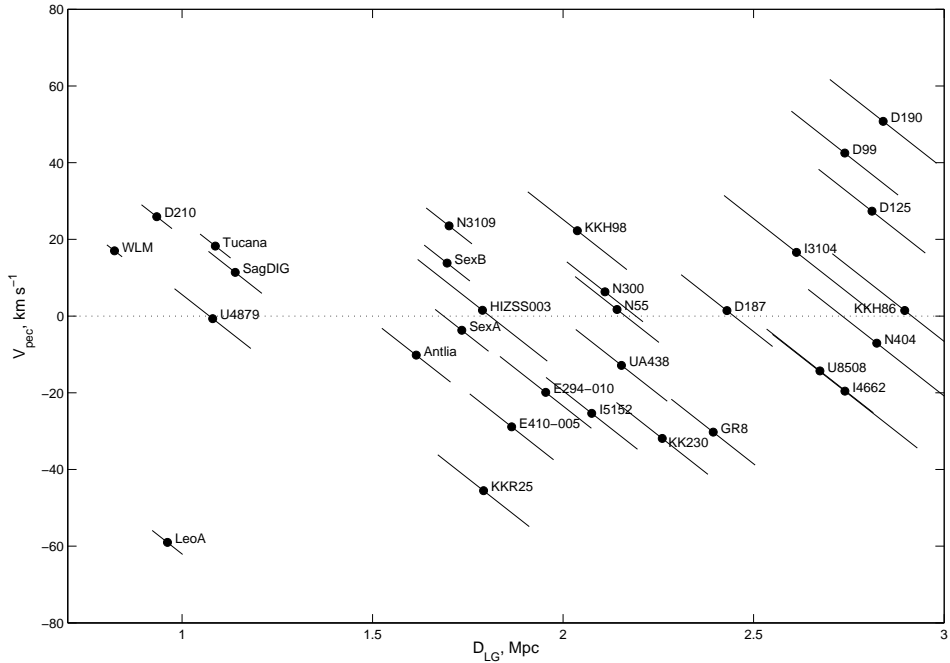}
\caption{Peculiar velocities and distances to the nearby galaxies with respect to the centroid of the LG. The inclined bars represent the measurement errors.}
\end{figure}

\begin{figure}
\includegraphics[scale=0.5]{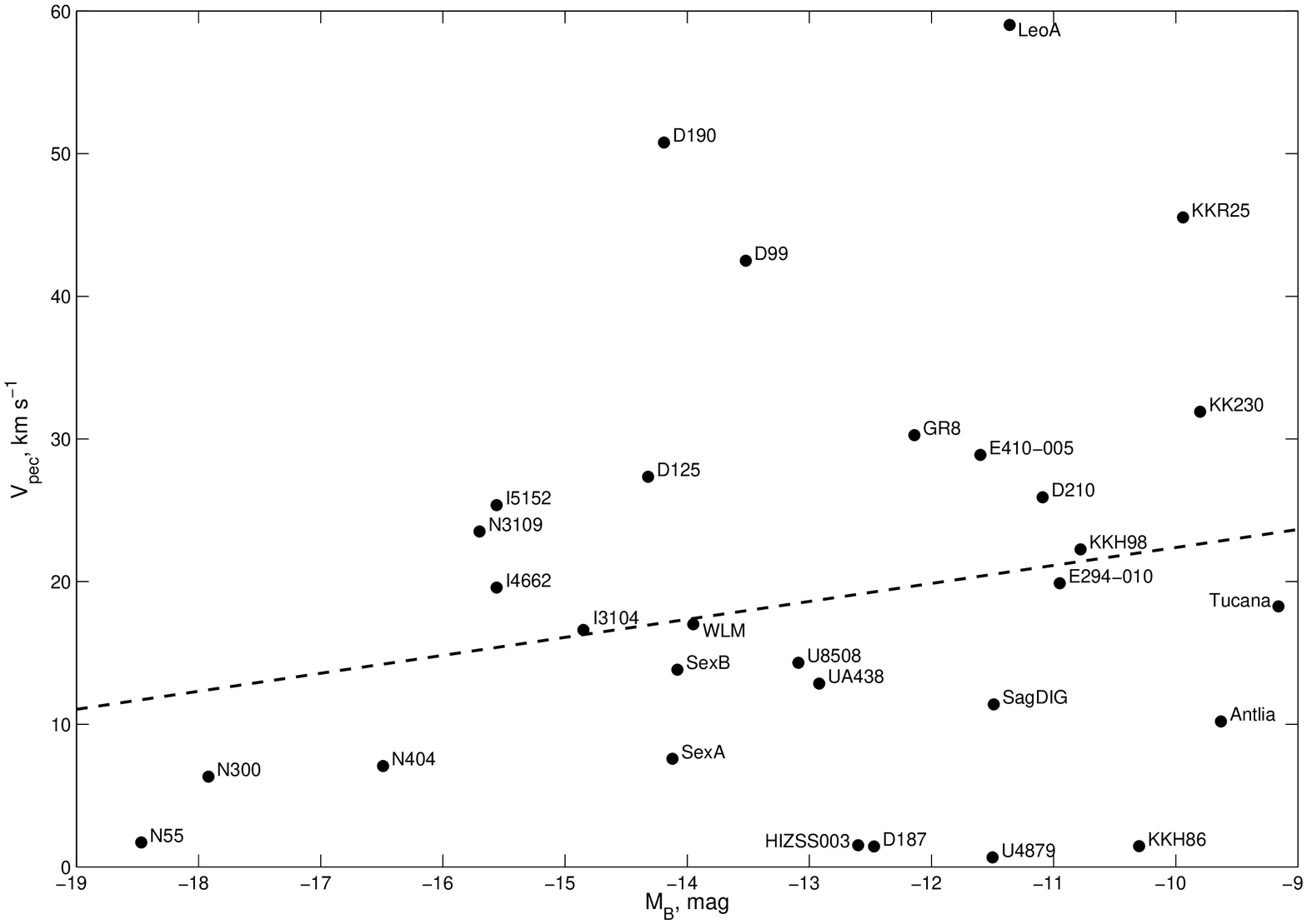}
\includegraphics[scale=0.5]{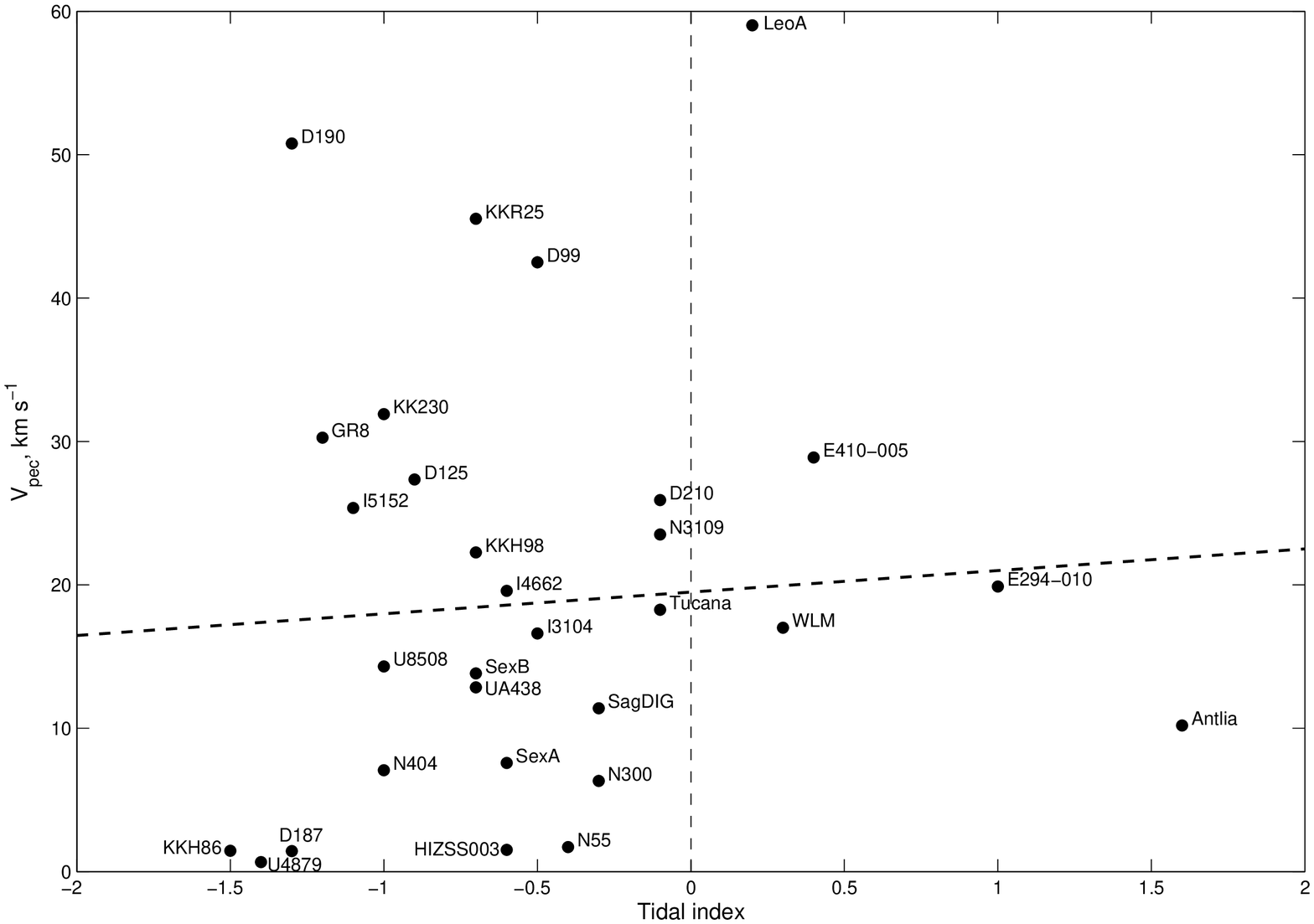}
\caption{Peculiar velocities of the nearby galaxies vs. their absolute magnitudes (upper panel) and vs. their tidal index (bottom panel).}
\end{figure}

\begin{figure}
\includegraphics[scale=0.5]{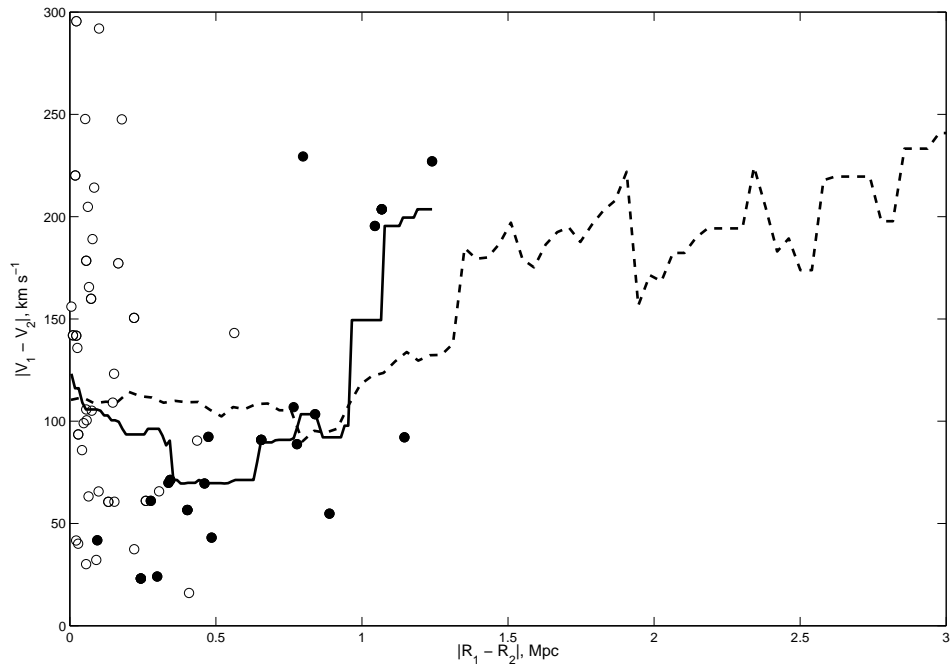}
\caption{Radial velocity difference for a galaxy and its nearest neighbour vs. their spatial separation. Open circles represent the LG members while filled ones correspond to the galaxies situated outside the virial region. Solid zigzag line is a sliding median for the LG surroundings, and the dashed one corresponds to a sliding median over the whole Local Volume (CNG).}
\end{figure}

\end{document}